\documentclass[preprint]{emulateapj}
\shorttitle{A hot horizontal branch star with a close K-type main-sequence companion}
\shortauthors{Moni Bidin et al.}

\begin{document}

\title{A hot horizontal branch star with a close K-type main-sequence companion}
\thanks{Based on observations with the ESO Very Large Telescope at Paranal
Observatory, Chile (program IDs 69.D-0682, 075.D-0492, and 079.D-0674), and with the 6.5m Magellan Telescopes at Las Campanas Observatory, Chile (program IDs
2006B-LC-7 and CHILE-2007B-007)}

\author{C. Moni Bidin\altaffilmark{1}, Y. Momany\altaffilmark{2}, M. Montalto\altaffilmark{3}, M. Catelan\altaffilmark{4,5},
S. Villanova\altaffilmark{6}, G. Piotto\altaffilmark{2,7}, and D. Geisler\altaffilmark{6}}

\altaffiltext{1}{Instituto de Astronom\'ia, Universidad Cat\'olica del Norte, Av. Angamos, 0610, Antofagasta, Chile; cmoni@ucn.cl}
\altaffiltext{2}{Istituto Nazionale di Astrofisica - Osservatorio Astronomico di Padova, Vicolo dell'Osservatorio 5, 35122 Padova, Italy}
\altaffiltext{3}{Instituto de Astrof\'isica y Ci\^{e}ncia del espa\c{c}o, Universidade do Porto, Rua das Estrelas, 4150-762, Porto, Portugal}
\altaffiltext{4}{Instituto de Astrof\'isica, Pontificia Universidad Cat\'olica de Chile, Casilla 306, Santiago, Chile}
\altaffiltext{5}{Millenium Institute of Astrophysics, Santiago, Chile}
\altaffiltext{6}{Departamento de Astronom\'ia, Universidad de Concepci\'on, Casilla 160-C, Concepci\'on, Chile}
\altaffiltext{7}{Dipartimento di Fisica e Astronomia ``Galileo Galilei'', Universit\'a di Padova, Vicolo dell'Osservatorio 3, Padova, I-35122 Padova, Italy}

\begin{abstract}
Dynamical interactions in binary systems are thought to play a major role in the formation of extreme horizontal branch stars (EHBs) in the Galactic field.
However, it is still unclear if the same mechanisms are at work in globular clusters, where EHBs are predominantly single stars. Here we report on the
discovery of a unique close binary system (period $\sim$1.61~days) in the globular cluster NGC\,6752, comprising an EHB and a main-sequence companion of
0.63$\pm$0.05~M$_\odot$. Such a system has no counterpart among nearly two hundred known EHB binaries in the Galactic field. Its discovery suggests that
either field studies are incomplete, missing this type of systems possibly because of selection effects, or that a particular EHB formation mechanism is
active in clusters but not in the field.
\end{abstract}


%%%%%%%%%%%%%%%%%%%%%%%%%%%%%%%%%%%%%%%%%
%%%%%%%%%%%%%%%%%%%%%%%%%%%%%%%%%%%%%%%%%

\section{Introduction}
\label{s_intro}

Extreme horizontal branch stars (EHBs) are hot (T$_\mathrm{eff}>$20\,000~K), evolved stars of low initial mass (0.7--2~M$_\odot$) burning helium in their
core \citep{Faulkner72,Heber86}. They have lost most of the hydrogen envelope during their evolution, to the point that the external layer is too thin to
sustain the hydrogen burning shell. Thus, after the exhaustion of helium, they are expected to evolve directly to the white dwarf cooling sequence
\citep[``AGB-manqu\'e stars''][]{Greggio90}. They are found both in the Galactic field and in globular clusters (GCs), and the comprehension of their
formation mechanisms is required to understand the late stages of the evolution of low-mass stars. They are also responsible for the ultraviolet emission of
elliptical galaxies \citep{Greggio90,Han07,Chung11}.

The causes of the heavy mass loss required to form an EHB star are still under debate, and this is one of the foremost grey points of low-mass stellar
evolution theory. Dynamical interactions inside binary systems are considered a major channel to produce an EHB star \citep{Han02,Han03}, as confirmed by the
high frequency of close binaries among EHBs in the Galactic field \citep[e.g.,][]{Kawka15}. However, it is unclear if the same mechanisms are at work in the
denser environment of GCs, where close EHB binaries are very rare \citep{Moni06,Moni09}. The old age of the cluster stellar population was proposed as an
explanation \citep{Moni08,Han08}, but an alternative scenario has become more popular in the last decade: EHBs could be the progeny of second-generation,
single, helium-enriched stars \citep{DAntona02,Chung11,Dalessandro11}.

The star M5865 in NGC\,6752 is so far the only spectroscopically-confirmed EHB close binary in a GC \citep{Moni08}. Later after discovery, \citet{Moni10}
showed that the presence of a G- or K-type main sequence (MS) companion had previously been proposed for this star by \citet{Moni07}, to explain its red color
and a faint MgIb triplet in its spectrum. A EHB star with a close companion of this kind has never been observed among the more than 180 EHB binaries discovered
to date in the Galactic field \citep{Ostensen06,Wade14}. In fact, their close companions are all either compact remnants (white dwarfs or neutron stars), or
very low-mass MS stars \citep[M$_\mathrm{MS}<$0.3~M$_\odot$, e.g.,][]{MoralesRueda03,Maxted01}. More massive MS companions have been found only in wide pairs
\citep[e.g.,][]{Vos13} or hierarchical triple systems \citep{Heber02}. Indeed, M5865 could be such a rare triple system, where the MS star in a wide orbit does
not interfere with the evolution of the inner, close pair. Alternatively, the MS star could also be a foreground object physically unrelated
to the EHB star, a common fact in the crowded environment of GCs.

We analyzed the available data of M5865 to identify the role of the MS star in the system. The data sets consisted of $i$) twenty-one high-resolution spectra;
$ii$) five intermediate-resolution spectra; $iii$) composite ground-based optical images; $iv$) more than 800 frames of time-series photometry in the V band.

%%%%%%%%%%%%%%%%%%%%%%%%%%%%%%%%%%%%%%%%%
%%%%%%%%%%%%%%%%%%%%%%%%%%%%%%%%%%%%%%%%%

\section{Data analysis}
\label{s_data}

\subsection{The mass of the MS star}
\label{ss_mass}

\begin{figure}
\includegraphics[angle=-90,scale=.9]{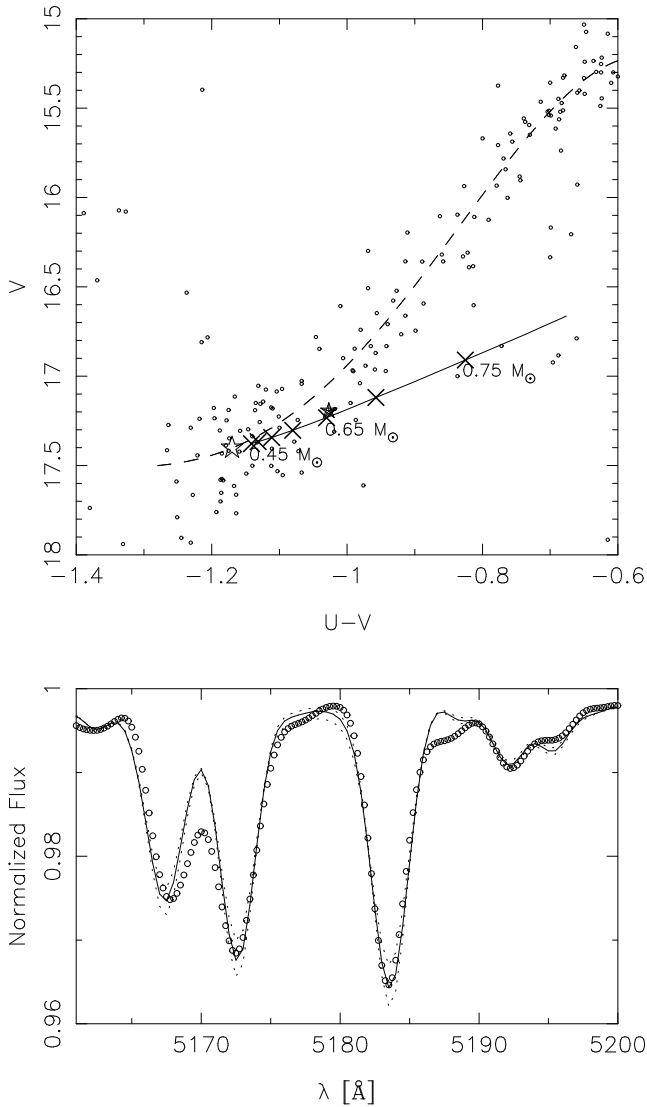}
\caption{{\it Upper panel}: Photometric mass estimate for the MS star. The filled and empty stars indicate the observed and theoretical (as single star)
position of M5865, respectively. Other cluster stars are shown with small circles, and the dashed curve indicates the polynomial fit of the cluster HB.
Solid curve and crosses show the positions of a EHB star plus a MS star of increasing mass, from 0.45 to 0.78~M$_\odot$. {\it Lower panel}: Spectroscopic mass
estimate for the MS star. The observed MgIb triplet (empty circles) is compared to synthetic spectra of a EHB star with a MS companion of varying mass. The
best-fit model (T$_\mathrm{eff,MS}$=5300~K) is indicated by the thick curve. The dashed curves show the same synthetic spectrum, with T$_\mathrm{eff,MS}$ varied
by $\pm$100~K. \label{f_mass}}
\end{figure}

We estimated the mass of the MS star (M$_\mathrm{MS}$), comparing the position of the star in the color-magnitude diagram (CMD) with synthetic loci
representative of a EHB star plus a MS companion. We empirically determined the starting point of the curve, i.e. the position of a single cluster HB star
($(U-V)_\mathrm{EHB}$, $V_\mathrm{EHB}$) at the temperature of the hot component of the system \citep[T$_\mathrm{eff,MS}=27800\pm300$~K,][]{Moni07}. We fitted
a $(U-V)$--T$_\mathrm{eff}$ relation to the hotter part of the horizontal branch (HB) of NGC\,6752, matching photometric colors
\citep[obtained with WFI at the MPG/ESO Telescope,][]{Momany02} with
spectroscopic temperature measurements \citep{Moni07}. We excluded all the stars with ``anomalous'' spectroscopic results, in particular the eight EHBs with
high spectroscopic mass, and we were left with 23 stars. We thus derived $(U-V)_\mathrm{EHB}=-1.18\pm0.03$. The error was estimated from the scatter of the
points with respect to the fitted curve, as also done for the magnitude obtained in the next step. We then fitted a fourth-order polynomial to the HB in the
CMD (dashed curve in Fig.\ref{f_mass}), and from the color $(U-V)_\mathrm{EHB}$ we derived $V_\mathrm{EHB}=17.39\pm0.18$. Given the expected photometric
properties of the EHB component alone, we calculated the resulting magnitude and color when the flux of a MS star between M$_\mathrm{MS}$=0.45 and
0.78~M$_\odot$ is added. This was taken from Yale-Yonsei theoretical isochrones \citep{Spada13} for metallicity Z=0.0004 and age=12.5 Gyr \citep{Vandenberg13},
assuming $(m-M)$=13.38, E($B-V$)=0.04 \citep{Harris96}, and a standard reddening law \citep{Cardelli89}. We thus obtained the curve shown in Fig.~\ref{f_mass},
with color and magnitude increasing with M$_\mathrm{MS}$. The point closest to the observed position returned M$_\mathrm{MS}=0.64\pm0.07$M$_\odot$. The
error was estimated from the variations of the result when the starting point was shifted by 1$\sigma$ in either magnitude or color (assuming an error of
0.05~mag in distance modulus). The photometric error of the observed position was considered also, but it resulted negligible (0.006 and 0.012~mag in $V$ and
($U-V$), respectively).

The MgIb triplet, signature of the MS star, was observed in five intermediate-resolution (R=4000) spectra collected with FORS2 at Paranal Observatory between
2002, June~11 and 14, with the 1400V grism and a 0.5$\arcsec$-wide slit. We remand the reader to \citet{Moni06} for more details about the observations and the
data reduction. We obtained a spectroscopic estimate of M$_\mathrm{MS}$ by fitting this feature with synthetic spectra. We shifted the spectra to laboratory
wavelength and co-added them. Flux-calibrated synthetic spectra were calculated with the SPECTRUM LTE spectral synthesis code \citep{Gray94}, fed with model
atmospheres obtained by interpolating on the \citet{Kurucz93} grid. The temperature and gravity of the EHB star were taken from our previous measurements
\citep{Moni07}. We calculated five models with T$_\mathrm{eff,MS}$ between 3800 and 5800~K, in steps of 500~K. Their gravities and masses were taken from the
same models used before. We interpolated the three hottest models at a step of 100~K to produce a finer grid in the range T$_\mathrm{eff,MS}$=4800--5800~K. The
stellar radii were estimated as $R=(MG/g)^{0.5}$, where $g$ is the surface gravity and $M$ is the mass, assuming 0.5~M$_\odot$ for the EHB star. We then scaled
the fluxes for the different radii of the two components, and finally co-added them. We compared each model with the observed spectrum, calculated the $\chi^2$
statistics, and fitted its dependence on T$_\mathrm{eff,MS}$ with a third-order polynomial. We thus found a minimum $\chi^2$ at T$_\mathrm{eff,MS}=5310\pm140$~K,
corresponding to an early K-type MS star with M$_\mathrm{MS}=0.62\pm0.03$~M$_\odot$. The fit is shown in Fig.~\ref{f_mass}. The errors were estimated from the
statistical behavior of the $\chi^2$ function. They are likely underestimated, because only random noise of the observed spectrum is taken into account, while
systematic effects, such as errors in the continuum placement or the exact magnesium abundance, can occur.

The two independent approaches returned very similar results. Accounting for possible systematics, we derive M$_\mathrm{MS}$=0.63$\pm$0.05~M$_\odot$. According
to the employed models, the cool star is therefore an early K-type MS object with T$_\mathrm{eff,MS}=5310\pm$20~K and $V_\mathrm{MS}=19.6\pm$0.1.

\subsection{Orbital solution of the EHB}
\label{ss_HR}

\begin{figure}
\includegraphics[angle=-90,scale=.43]{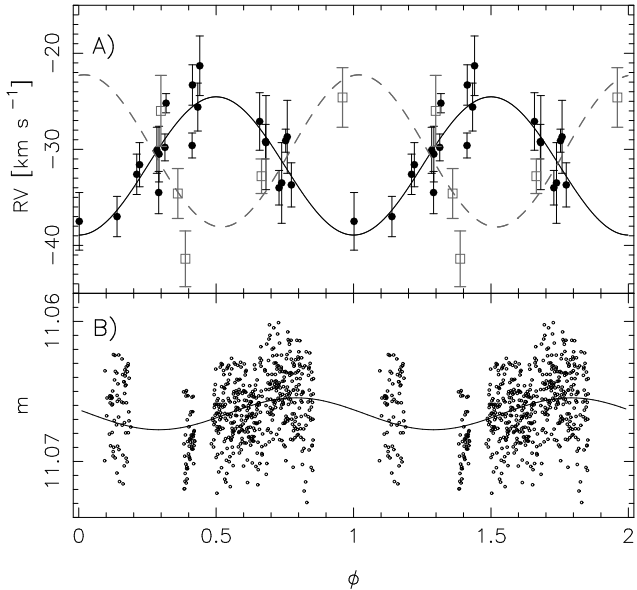}
\caption{{\it Upper panel}: Phased RV curve of M5865. The filled circles indicate the RV of the H$_\beta$ line measured on high-resolution spectra. The black
curve shows the orbital solution of the EHB component derived from these data. The gray squares indicate the RV of the MgIb triplet. These data were phased with
the ephemeris obtained for the EHB star, and the dashed, gray curve shows their best-fit sinusoidal curve. {\it Lower panel}: Photometric time series of M5865 (in
instrumental $V$ magnitudes), phased with the same ephemeris of upper panel. The best-fit sinusoidal is shown. \label{f_phase}}
\end{figure}

Seventeen high-resolution (R=18000) spectra centered on the H$_\beta$ line were collected between June and September 2007 in service mode at Paranal Observatory,
with the FLAMES-GIRAFFE spectrograph and the H7A setup. This data set was complemented with four more epochs, collected on 2005 June 29, with the same instrument
and setup \citep{Moni08}. We reduced the data by means of the CPL-based ESO pipeline. We checked the wavelength calibration analyzing the extracted and calibrated
lamp fibers, and we found only small random deviations from the laboratory wavelengths (0.3~km~s$^{-1}$ rms). We extracted the science spectra with an optimum
algorithm \citep{Horne86}, and subtracted the smoothed average of ten fibers allocated to the sky background. Some spectra presented a wide emission feature on
the red wing of the H$_\beta$ line, as a result of contamination from a nearby lamp fiber. We fitted it with a Gaussian profile and removed it. We repeated the
procedure varying the constraints (continuum definition and fitted data points), to estimate the uncertainty thus introduced on the RVs. This was negligible
($<1$~km~s$^{-1}$) in all but two spectra, where it was of the order of 2.5~km~s$^{-1}$. To measure the RVs, we cross-correlated \citep{Tonry79} the H$_\beta$
line with its wings (4840--4880~\AA) with a synthetic spectrum with parameters matching those of the EHB star, drawn from the library of \citet{Munari05}. We
corrected the observed RVs to heliocentric values. The spectra of about 60~hot stars were contemporarily observed by the multi-object GIRAFFE spectrograph. We
estimated and removed zero-point offsets between the frames forcing the average RV of the 40 best targets (in terms of spectral quality and measurement accuracy,
and showing no relevant RV variation) to be constant among the epochs \citep[see][for more details]{Moni11}. We estimated the associated uncertainty from the
related error-on-the-mean.

\begin{figure}
\includegraphics[angle=-90,scale=.35]{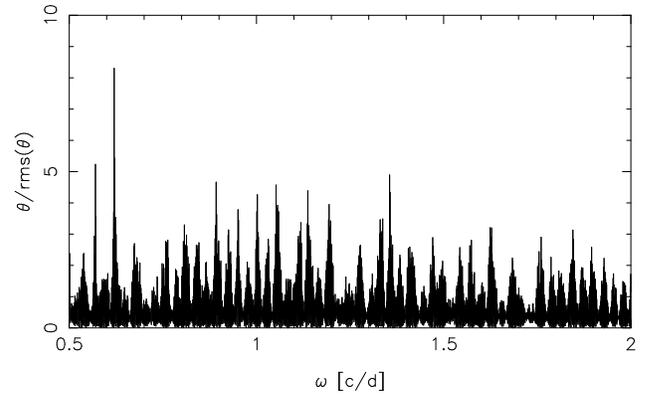}
\caption{ANOVA periodogram of the RV data. The power spectrum is given in units of the rms. \label{f_anova}}
\end{figure}

We analyzed the periodograms of the data calculated with several algorithms, such as the Analysis of Variance
\citep[ANOVA,][shown as an example in Fig.~\ref{f_anova}]{SchwarzenbergCzerny96}, the Lomb-Scargle algorithm \citep{Lomb76,Scargle82}, the \citet{Bloomfield76}
Fourier Analysis, the Data-Compensated Discrete Fourier Transform \citep{FerrazMello81}, and the Fourier Analysis of Light Curves of \citet{Harris89}. A prominent
peak at P=1.61~days dominates all the periodograms. The corresponding phased RV curve is shown in Fig.~\ref{f_phase}. This solution is robust, because it is
stable against the exclusion of up to five of the most uncertain measurements, and none of the secondary peaks returned an equally satisfactory solution, in terms
of rms deviation and $\chi^{2}$-statistics of the fit. Despite the high significance of the RV variations (4.6$\sigma$), the semiamplitude of the RV curve
(7.2$\pm$0.7~km~s$^{-1}$) is small compared to the typical variations of close EHB systems \citep{MoralesRueda03}. This suggests a nearly face-on orbital plane.

\subsection{Visual companion}
\label{ss_images}

Unfortunately, M5865 is outside the field of view of all HST archive images. We therefore stacked our thirteen best-seeing, ground-based, high-spatial resolution
frames in search of hints of a MS star projected along the line of sight. No physical companion should be detected, as even a wide binary system would be
unresolved at the cluster distance \citep[4.2~kpc,][]{Harris96}. The data were collected with MagIC at the Clay telescope at Las Campanas Observatory on May 9,
2006, with a spatial resolution of $\sim0.1\arcsec$ per pixel. The effective seeing of the composite image is 0.41$\arcsec$, and M5865 has exactly the same
full-width-at-half-maximum as other isolated stars in this
frame. Any blended star should be closer than 0.16$\arcsec$, or a larger point-spread-function would have been detected. M5865 is located at a distance to the
center of twice the cluster half-mass radius \citep{Harris96}, where the crowding conditions are not severe. The density of stars within 0.5$\arcmin$ from M5865
in the WFI photometric catalog \citep{Momany02}, corrected for the catalog's completeness (95\% for stars with $V<20.5$), is $\sim$0.028~stars~arcmin$^{-2}$.
Hence, the probability of a chance aligment with separation $<0.16\arcsec$ is 0.2\%. As discussed in \S\ref{ss_LR}, the MS star is RV variable. The fraction of
MS binary stars in the cluster is extremely low \citep[$<$1\% in the external regions,][]{Milone10}, and the joint probability of having a blended MS foreground
star that is also a binary is negligible ($<2\cdot 10^{-3}$). The MS star must therefore be part of the M5865 system.

\subsection{RV variations of the Mg triplet}
\label{ss_LR}

We used the FORS2 spectra presented in \S\ref{ss_mass} to measure and compare the RV of the H$_\beta$ line and of the MgIb feature. \citet{Moni06} showed that
these data are affected by RV zero-point offsets up to 10--15~km~s$^{-1}$, and they are suitable only for measurements of RV variations, with precision of
3--4~km~s$^{-1}$. We therefore measured the RV variations ($\Delta\mathrm{RV}_i$) in the spectra with respect to the first spectrum of the series, and we
corrected them as detailed in \citet{Moni06}. We found RV variations of the MgIb triplet, significant at the 3.8$\sigma$ level. We then derived the absolute RV
of the reference spectrum (RV$_\mathrm{ref}$), cross-correlating it with the best-fit synthetic template used for the spectroscopic estimate of M$_\mathrm{MS}$.
The combination of $\Delta\mathrm{RV}_i$ and RV$_\mathrm{ref}$ returned the absolute RV in each spectrum, but all these measurements were offset by the
zero-point systematic of the reference spectrum. Hence, we phased the RVs of the H$_\beta$ line with the ephemeris obtained in Sect.\ref{ss_HR}, then we
compared them with the best-fit RV curve (black curve in Fig.~\ref{f_phase}), and we minimized the $\chi^2$ statistics to find the zero-point correction.

The third object of a hierarchical triple system must show a low-amplitude, low-frequency RV curve unrelated to the orbital period of the inner pair. The RV
variations of the MgIb triplet are not compatible with this scenario, because they are too large (comparable to those of the EHB star) in a temporal interval of
only 3~days, and they are in rough antiphase with the variations of the H$_\beta$ line on the same spectra. Unfortunately, the latter are not significant when
compared to the large errors \citep{Moni06}, and the measurements are too few to independently derive the periodicity of the MS star. However, when we phase
the RV of the MS star with the solution previously obtained from the high-resolution spectra, we find that the best-fit sinusoidal curve is in antiphase with
that of the EHB star, as shown in Fig.\ref{f_phase}. Thus, the behavior of the MgIb triplet demonstrates that the MS star is the close companion of the hotter
component.

\subsection{Light variations}
\label{ss_lightcurve}

\begin{figure}
\includegraphics[angle=-90,scale=.4]{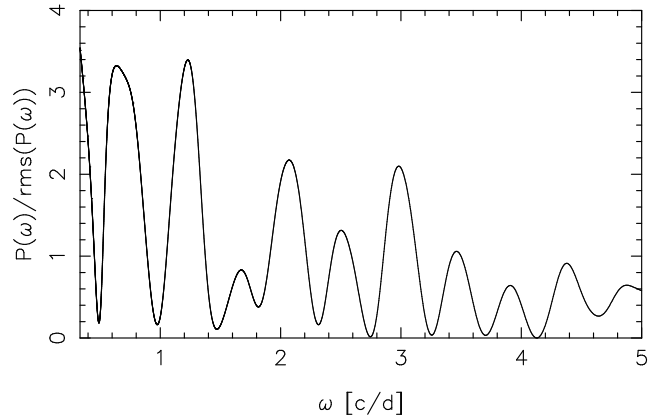}
\caption{Lomb-Scargle periodogram of our time series photometry of M5865. The power spectrum is given in units of the rms scatter. \label{f_imacs}}
\end{figure}

An EHB star with a close MS companion should show light variability \citep{Heber04,Geier12}. Stars deformed by tidal forces show ellipsoidal variations
faster than the orbital period by a factor of two, with maxima at the phases of maximum and minimum RV. On the other hand, a star in rotationally locked orbit
exhibits reflection effects on the heated hemisphere, synchronous with the orbital period and with maximum brightness at phase $\Delta\phi$=0.25 before the
minimum RV.

More than 800 frames in the $V$ band were acquired with IMACS at the 6.5m Baade telescope at Las Campanas Observatory, during three consecutive nights
\citep[2007 July 7 to 9,][]{Catelan08}. The instrument was used at f/4.3 and binned 2$\times$2 on the chip, for a resulting scale of 0.22$\arcsec$ per pixel.
Exposure times varied between 30s and 40s. The weather was good during the first and third night, with average seeing of 0.7$\arcsec$ and 1.1$\arcsec$,
respectively. The second night was affected by strong wind, and the seeing varied between 1$\arcsec$ and 1.8$\arcsec$. We performed the photometry using
standard image subtraction routines as implemented in ISIS2 \citep{Alard98,Alard00}. We first constructed a high S/N (reference) image from the 50 best-seeing
images. This was convolved with a kernel to match the seeing of each input frame, with a background variable at the second order, and subtracted to it after
astrometric alignment. Aperture photometry was then performed on the difference image by means of our own customized software. We masked out bad or saturated
pixels, and all stars within 20 pixels from such bad features were not further considered. We fixed the aperture radius to the Full-Width-at-Half-Maximum of
the input image, and the inner and outer radius of the sky annulus to 20 and 35 pixels, respectively. We then subtracted from the light curves the median
photometric zero point between each frame and the first one of the series. We considered only the brightest stars in its calculation, and we applied a
3$\sigma$-clipping algorithm to exclude outliers. We eventually removed a tiny residual trend, of the order of a few mmag, forcing the average instrumental
magnitude of 43~stars within 1$\arcmin$ from the target to be constant among the frames. We also applied a 3$\sigma$-clipping selection to clean the light
curve from outliers, and excluded the data points affected by large errors. We thus ended up with 585 measurements spanning a temporal interval of about
2.5~days.

The resulting periodogram is shown in Fig.\ref{f_imacs}. It is characterized by wide and poorly-significant peaks, as a consequence of the poor temporal
sampling. However, the two dominant peaks match extremely well the orbital period and its first overtone (P$\sim$1.62 and 0.81~days). The corresponding light
variation is tiny (of the order of 2~mmag peak-to-peak), a fact is consistent with a high inclination of the orbital plane along the line of sight. The
position of the maximum at phase $\phi$=0.75, as shown in Fig.~\ref{f_phase}, excludes ellipsoidal variations with P$\sim$0.81~days, and it suggests that the
dominant variability is due to reflection effects (P$\sim$1.61~days) on the surface of the EHB star. This fact is new, because in known systems the MS object
is too cold, or too far away, to noticeably heat the EHB companion. Clearly, better time series are required to study the irradiation from this star in more
detail.

%%%%%%%%%%%%%%%%%%%%%%%%%%%%%%%%%%%%%%%%%
%%%%%%%%%%%%%%%%%%%%%%%%%%%%%%%%%%%%%%%%%

\section{Conclusions}
\label{s_conclusions}

Our study reveals that the close companion of the EHB star M5865 is a MS star of mass M$_\mathrm{MS}\approx 0.6$M$_\odot$. This discovery does not
increase the low estimate of the close binary fraction in the cluster \citep[$\sim$4\%,][]{Moni08}, because this binary had already been
identified. However, it marks a striking difference with field studies, where close binaries are very common but such a system has never been observed.

The theoretical models predict the formation of few such objects, resulting from a first common-envelope (CE) stage, although low-mass companions are highly
preferred \citep{Han03,Yungelson05}. In an old stellar population, these systems could be more common than EHBs with a compact companion, products of a
second-CE phase \citep{Han08}. Hence, it could not be completely surprising that the first EHB close binary studied in a GC is such a unique pair.
Nevertheless, their absence in the field is harder to explain because, while \citet{Han03} shows that first-CE products could be a factor of two less frequent
at higher metallicity, \citet{Han08} finds no clear decrease of the first-CE channel efficiency at younger ages. M5865-like systems have so far been
overlooked in the literature due to the lack of observational counterparts, and theoretical models should study their formation mechanism in more details.
It is also possible that they have remained undetected in the field due to selection effects such as the known ``GK'' bias \citep{Han03}.
In this case, the discovery of M5865 would indicate that the studies of field EHB stars are incomplete, because a class of systems has so far been neglected.

%%%%%%%%%%%%%%%%%%%%%%%%%%%%%%%%%%%%%%%%%
%%%%%%%%%%%%%%%%%%%%%%%%%%%%%%%%%%%%%%%%%

\acknowledgments
CMB, SV, and MC acknowledge support from proyectos FONDECYT regular 1150060, 1130721, and 1141141. MC acknowledges support by the Ministry of Economy,
Development, and Tourism's Programa Iniciativa Cient\'ifica Milenio, through grant IC210009, awarded to the Millenium Institute of Astrophysics (MAS) by
Proyecto Basal PFB-06/2007. MM acknowledges support from FCT through the grant and SFRH/BDP/71230/2010.

%\plottwo{f2.eps}{f2_color.eps}
%\includegraphics[angle=90,scale=.50]{f3.eps}

\end{document}